\documentclass{article} %
\usepackage{iclr2022_conference,times}

\usepackage[utf8]{inputenc}
\usepackage{amsmath}
\usepackage{amssymb}
\usepackage{amsthm}
\usepackage{amsfonts}
\usepackage{bm}
\usepackage{booktabs}
\usepackage{algorithm}
\usepackage{algpseudocode}
\newcommand*\Let[2]{\State #1 $\gets$ #2}
\usepackage{microtype}
\usepackage{graphicx}
\usepackage{xcolor}
\usepackage{url}

\usepackage{hyperref}
\hypersetup{
    colorlinks=true,
    linkcolor={blue!50!black},
    citecolor={blue!50!black},
    urlcolor={blue!80!black}
}

\def\eqref#1{equation~\ref{#1}}

\def\vs{{\bm{s}}}

\DeclareMathAlphabet{\mathsfit}{\encodingdefault}{\sfdefault}{m}{sl}
\SetMathAlphabet{\mathsfit}{bold}{\encodingdefault}{\sfdefault}{bx}{n}

\newcommand{\V}[1]{\bm{#1}}

\newcommand{\eq}[1]{\begin{align}#1\end{align}}

\newcommand{\brck}[1]{\left(#1\right)}

\newcommand{\brcksq}[1]{\left[#1\right]}
\newcommand{\brckcur}[1]{\left\{#1\right\}}

\def\shownotes{1}  %
\ifnum\shownotes=1
\newcommand{\authnote}[2]{{\scriptsize $\ll$\textsf{#1 notes: #2}$\gg$}}
\else
\newcommand{\authnote}[2]{}
\fi
\ifnum\shownotes=1

\else

\fi

\def\DGEmodel{DGE}
\def\RBCmodel{RBC}
\def\idxi{i}

\def\idxj{j}
\def\NumConsumers{|J|}

\def\idxplanner{p}
\def\xbudgetb{  {\color{black} B}}
\def\xtryconsumptionc{  {\color{black} \hat{c}}}
\def\xconsumptionc{ {\color{black} c}}
\def\xtotconsumptionC{  {\color{black} C}}
\def\xcapinvestk{   {\color{black} \Delta k}}
\def\xcapitalk{ {\color{black} k}}
\def\xtotcapitalK{  {\color{black} K}}
\def\xhoursn{   {\color{black} l}}
\def\xtothoursN{    {\color{black} L}}
\def\xobso{ {\color{black} o}}
\def\xpricep{   {\color{black} p}}
\def\xprofitP{  {\color{black} P}}
\def\xquantityq{    {\color{black} q}}

\def\xinventoryS{   {\color{black} y}}
\def\xwagew{    {\color{black} w}}
\def\xproductionY{  {\color{black} Y}}
\def\prodfuncfactor{    {\color{black} A}}
\def\prodfuncalpha{\alpha}
\def\xtimestept{t}
\def\income{z}
\def\xincomeI{  {\color{black} \income}}
\def\incometaxrate{ {\tau } }
\def\corptaxrate{ {\sigma } }
\def\xtaxrevR{  {\color{black} R}}

\def\laborcoeff{\theta}

\def\nnfeat{\varphi}

\def\consumersymbol{c}
\def\firmsymbol{f}

\def\governmentsymbol{p}

\def\socialwelfare{\textit{swf}}

\def\SWF{\socialwelfare}

\def\util{u}

\newcommand{\policy}{\pi}

\def\ob{o}

\def\ac{a}
\def\St{\vs}
\def\st{s}

\def\rew{r}

\def\Val{V}

\def\df{\gamma}
\def\pol{\pi}

\def\eplen{T}

\title{%
Analyzing Micro-Founded General Equilibrium Models with Many Agents using Deep Reinforcement Learning
}

\author{%
Michael Curry\thanks{Research conducted while Michael Curry was an intern at Salesforce.}\\
University of Maryland \\
\texttt{curry@cs.umd.edu} \\
\AND
Alexander Trott, Soham Phade, Yu Bai, Stephan Zheng \\
Salesforce \\
Palo Alto, CA \\
\texttt{(atrott,sphade,yu.bai,stephan.zheng)@salesforce.com}
}
\iclrfinalcopy
\date{\today}

\begin{document}
\maketitle
\begin{abstract}
Real economies can be modeled as a sequential imperfect-information game with many heterogeneous agents, such as consumers, firms, and governments. 
Dynamic general equilibrium (\DGEmodel{}) models are often used for macroeconomic analysis in this setting.
However, finding general equilibria is challenging using existing theoretical or computational methods, especially when using \emph{microfoundations} to model individual agents. 
Here, we show how to use deep multi-agent reinforcement learning (MARL) to find $\epsilon$-meta-equilibria over agent types in microfounded \DGEmodel{} models. 
Whereas standard MARL fails to learn non-trivial solutions, our structured learning curricula enable stable convergence to meaningful solutions. 
Conceptually, our approach is more flexible and does not need unrealistic assumptions, e.g., continuous market clearing, that are commonly used for analytical tractability.
Furthermore, our end-to-end GPU implementation enables fast real-time convergence with a large number of RL economic agents.
We showcase our approach in open and closed real-business-cycle (\RBCmodel{}) models with 100 worker-consumers, 10 firms, and a social planner who taxes and redistributes.
We validate the learned solutions are $\epsilon$-meta-equilibria through best-response analyses, show that they align with economic intuitions, and show our approach can learn a spectrum of qualitatively distinct $\epsilon$-meta-equilibria in open \RBCmodel{} models. 
As such, we show that hardware-accelerated MARL is a promising framework for modeling the complexity of economies based on microfoundations.
\end{abstract}

\let\svthefootnote\thefootnote
\newcommand\freefootnote[1]{%
  \let\thefootnote\relax%
  \footnotetext{#1}%
  \let\thefootnote\svthefootnote%
}

\section{
    Introduction
}\label{Section:Introduction}
Real-world economies can be modeled as general-sum sequential imperfect-information games with many heterogeneous agents \citep{mascolell_microeconomic_1995}, such as consumer-workers, firms, and governments (or other social planners).
Dynamic general equilibrium models (\DGEmodel{}) are workhorse models that describe the economic incentives, interactions, and constraints of these agents, which are often assumed to be rational.%
\footnote{%
One can also model more human-like behavior through instances of \emph{bounded rationality}, e.g., agents that act suboptimally or whose affordances or objectives encode cognitive biases or limits \citep{kahneman2003maps}.
In this work, we focus on rational agents, but our framework can be generalized to using boundedly rational agents.
}
In particular, we are interested in \DGEmodel{} models \emph{with a large number of heterogeneous, strategic agents}, based on appropriate \emph{microfoundations} \cite{archibald1970microeconomic,smets2007shocks}.

By finding the strategic equilibria in such games, one can study \emph{macroeconomic} outcomes, such as productivity, equality, and growth \citep{heer2009dynamic}.
However, at this scale and game-theoretic complexity, existing theoretical and computational methods often struggle to find the strategic equilibria \cite{nisan_roughgarden_tardos_vazirani_2007,bai2021sampleefficient}.
For a detailed exposition on \DGEmodel{} models and solution methods related to our work, see Section \ref{Section:Related Work}.
We emphasize that we focus on the methodological challenge of finding equilibria, rather than the question of what constitute appropriate microfoundations.

In this work, we focus on reinforcement learning as a powerful and flexible methodological framework to analyze \DGEmodel{} models \emph{with many agents}.
We propose using deep \emph{multi-agent reinforcement learning (MARL)} \cite{sutton2018reinforcement} as a constructive solution to explicitly find their (approximate) equilibria. 
Using MARL provides many benefits:
1) large-scale deep RL has proven capable of finding (near-)optimal behavioral policies in complex multi-agent games in various domains \cite{alphastarblog,OpenAI_dota,silver2017mastering};
2) RL is flexible: it has few analytical requirements on the structure of the game, e.g., it can optimize policies for any scalar objective, e.g., consumer utility function or social welfare, which does not need to be differentiable; and 
3) it can optimize rich behavioral models, e.g., deep neural networks, that can imitate complex human behaviors, e.g., given multi-agent behavioral data \cite{zheng2016generating,zhan2018generative}.
As such, deep MARL holds promise as a framework to model macroeconomic outcomes based on microfoundations.

\paragraph{Economic Interactions Pose Learning Challenges.}
Although RL offers many conceptual benefits for economic modeling, training each agent in a \DGEmodel{} model independently using RL often fails due to economic interactions.
This is because economic interactions introduce \emph{mechanism design} problems between agents. 
For example, firms choose how to set prices when interacting with a population of consumers. 
Furthermore, a consumer's purchasing power changes when firms change wages or governments change taxes.

As such, economic interactions imply that \emph{the actions of one economic agent can dramatically change the reward function and constraints (constituting a mechanism)} of other agents. 
Because RL agents learn through \emph{exploration}, agents sample (sub)-optimal policies during training, which may overly distort the rewards or constraints of other RL agents. 
As a result, RL agents that learn independently (and do not account for the learning process of other agents) often fail to learn non-trivial behaviors in \DGEmodel{} models. 
This non-stationary learning dynamic becomes especially challenging for \DGEmodel{} models with a large number of heterogeneous RL agents.

Moreover, \emph{which} equilibrium agents converge to (``equilibrium selection'') may depend on, e.g., the 
1) world dynamics, 
2) initial conditions and policies,
3) learning algorithm, and
4) policy model class. 
Our framework allows us to study how these factors relate to which equilibria can be found.

\subsection{Our Contributions}
To address these challenges, we show how to effectively apply MARL to macroeconomic analysis:
\begin{enumerate}
    \item To enable stable convergence of MARL with many agents, we \emph{generalize the MARL curriculum learning method} from \citet{zheng2020ai} to structured curricula that train multiple agent types across multiple phases. 
    This approach yields non-trivial solutions \emph{more stably} compared to using independent training of RL agents, which often fails. 
    We also show that an RL social planner can improve social welfare vs fixed baselines.
    \item We show that our MARL approach can explicitly find local $\epsilon$-equilibrium strategies for the meta-game over agent types, without approximating the \DGEmodel{} model. 
    Previous solutions only could do so implicitly or for approximations of the \DGEmodel{} dynamics. 
    Here, the meta-game equilibrium is a set of agent policies such that no \emph{agent type} can unilaterally improve its reward by more than $\epsilon$ (its \emph{best-response}). 
    Furthermore, we learn a \emph{spectrum of solutions} in open \RBCmodel{} economies, 
    \item Our approach yields \emph{new economic insights}. 
    Our approach is more flexible and can find stable solutions in a variation of real-business-cycle (\RBCmodel{}) models~\citep{PIERREDANTHINE19931}, a family of \DGEmodel{} models. 
    We find solutions in closed and open \RBCmodel{} economies, the latter has a price-taking export market.
    After training, the behavior of our agents shows sensible economic behavior -- for example, negative correlations of prices with consumption, and positive correlation of wages with hours worked, over a range of possible outcomes.
    Within each outcome, firms and consumers display different strategies in response to their conditions -- for example, firms with capital-intensive production functions invest in more capital.
\end{enumerate}

\paragraph{Enabling Economic Analysis with Many RL Agents.}
We are particularly interested in \DGEmodel{} models with a large number of \emph{heterogeneous strategic} agents. 
To enable MARL at this scale, we ran both simulation and RL training on a GPU using the WarpDrive framework \citep{lan2021warpdrive}. 
WarpDrive accelerates MARL by orders of magnitude, e.g., by avoiding copying data unnecessarily between CPU and GPU.
In effect, this enables MARL to converge in hours (rather than days). 
Such system design and implementation choices are key to enable richer economic analysis, without resorting to approximations like representative agents \cite{kirman1992whom,hartley2002representative} or mean-field methods \cite{yang2018mean}.
In contrast, prior computational work often was limited to training a small number of independent, strategic agents.

Using hardware acceleration to train machine learning models at larger scales has repeatedly resulted in qualitatively new capabilities in machine learning, including some of the most spectacular results in language modeling~\cite{brown2020language} and computer vision~\cite{krizhevsky2012imagenet} in the past decade.
An essential part of each of those breakthroughs was optimizing system design and implementation.
While we don't claim to match the impact of these results, we want to emphasize that we see improved engineering to enable economic modeling at greater complexity and scale as a key contribution.
In the context of economic modeling, being able to learn and analyze across a spectrum of possible outcomes and solutions is key for policymaking~\cite{haldane2019drawing}.

\paragraph{Code.}
The code for this paper will be publicly available for reproducibility.
\section{
    Prior Work
}\label{Section:Related Work}

We now discuss how our MARL approach contrasts with or complements prior work.

\subsection{The Economics of \DGEmodel{} Models}
\DGEmodel{} models study the behaviors of and interactions between consumers, firms, and perhaps a government.
The markets for goods, labor, and capital are often assumed to be competitive, with prices set to clear the markets at each time step.
Typically, consumers may work, save, and consume, and balance these choices to maximize their total time-discounted future utility.
In other words, their decisions represent the solution to a Bellman equation.
There are \DGEmodel{} models without~\cite{smets2007shocks} or with~\cite{kaplan2018monetary} heterogeneity among agents.

Variations of \DGEmodel{} models also consider \emph{open} economies with trade and flows with a (large) external economy; accordingly, closed economies lack such an external connection. 
Typically, they have the same structure of market-clearing for wages, prices, and capital, but also allow consumers to invest in foreign capital~\cite{mendoza1991real,de2019approximately}.
We consider both closed and open economies: we choose to model the open case by giving firms access to an export market with an \emph{inelastic} price for goods.

\subsection{Macroeconomic Modeling and Microfoundations}

\paragraph{Microfoundations.}
\citet{lucas1981after} argued for an approach to macroeconomic modeling based on microfoundations \cite{smets2007shocks}.
Broadly construed, microfoundations analyze macroeconomics through models with individual self-interested strategic agents; macroeconomic outcomes are then the aggregation of their behaviors.
In particular, microfoundations address the Lucas critique, which points out the issue of basing macroeconomic policy on historical data.
That is, historically observed aggregate patterns of behavioral responses to changes in economic policy may not extrapolate to the future, as individuals continuously adapt their behavior to the new policy.

\paragraph{Conceptual Limitations of Prior \DGEmodel{} models.}
\citet{stiglitz2018modern} discusses several limitations of previous \DGEmodel{} models, including the use of inappropriate microfoundations.
For instance, agents are often assumed to be rational, have stylized decreasing marginal utilities, 
and maximize their expected value in the presence of uncertainty (\emph{the rational expectations assumption}), in the face of perfectly clearing markets.
However, modern \DGEmodel{}s often lack agent models based on behavioral economics, for example.
We emphasize that our use of MARL is agnostic to the specifications of the individual agents, which may be rational or boundedly rational.

Moreover, while \DGEmodel{} models are based on microeconomic theory for individuals, they often do not actually model the large number of agents as individuals.
Rather, many models use \emph{representative agents} as a proxy for the aggregate behavior of, e.g., all consumers.  
These and other simplifying, but restrictive assumptions make it possible to find analytic solutions, but lead to unrealistic outcomes, e.g., no trading. 
Moreover, the use of representative agents does not lead to uniqueness and stability of equilibria \cite{kirman1992whom}, a popular desideratum.

\paragraph{Challenges of Solving \DGEmodel{} Models.}
Existing analytical and computational methods often struggle to find explicit \DGEmodel{} equilibria, as \DGEmodel{} models are typically highly nonlinear.
More generally, enumerating and selecting equilibria in general-sum games is an unsolved challenge \citep{bai2021sampleefficient}.

Analytical work often simplifies the \DGEmodel{} model's dynamics \cite{lucas1981after}, e.g., through linearization.
However, a linearized \DGEmodel{} model may have fewer or different equilibria compared to the full \DGEmodel{} model, and a linearized solution may prefer one equilibrium arbitrarily when many are possible. 
Furthermore, linearization may only be a good approximation around steady-state equilibria and not be valid in the presence of large shocks, e.g., when agents change their behavior dramatically in time \citep{stiglitz2018modern,BONEVA2016216,atolia2010misleading}.

Formal methods, e.g., backwards induction and other dynamic programming methods, often don't yield explicit solutions and only implicitly characterize optimal policies \citep{10.2307/j.ctvjnrt76}.
For example, in taxation, one can analyze the distortion in consumer savings or capital, the asymptotic behavior of optimal taxes in the far future, or the tax rate on the highest-skill agent \citep{golosov_optimal_2011,acemoglu2010dynamic}.
However, these methods generally cannot find the full tax policy explicitly or only describe part of it.

\paragraph{Computational Methods and Agent-Based Models.}
A related body of work has studied agent-based models (ABM)~\cite{bonabeau_agent-based_2002,sinitskaya2015macroeconomies,haldane2019drawing}.
ABMs may include a much larger range of heterogeneous agents (compared to typical \DGEmodel{} models), with possibly similar microfoundations, but whose affordances and dynamics are often highly stylized.
Through repeated simulation, ABMs typically study the emergent (macroeconomic) phenomena given simple behavioral rules for the agents, across a range of parameters. 
However, these studies mostly do not address questions of finding the equilibria and optimal behaviors.
Moreover, computational methods more generally still pose technical and conceptual challenges for economic analysis \cite{judd1997computational}. 
Libraries exist to use numerical methods, e.g. to find fixed-point solutions ~\cite{holden2017existence,mendoza2020fipit}.
However, it can be hard to distill simple or explainable economic principles from numerically-found optimal behaviors and simulations.

\subsection{Reinforcement Learning and Economics}
\paragraph{Modeling Economic Agents using RL.}
In the economics literature, \citet{haldane2019drawing} observed that RL is a natural fit for economic modeling.
Rational economic agents maximize their total discounted future reward, which is equivalent to the definition of an RL agent.
Their optimal behavioral policy (or simply ``policy'') are characterized by the Bellman equation. 
The economics literature traditionally solves such Bellman equations with methods that are only usable in fully-known environments, e.g., value or policy function iteration~\cite{coleman1990solving,judd1992projection}

In contrast, in the machine learning literature, many gradient-based RL techniques, e.g., REINFORCE \cite{Williams:1992:SSG:139611.139614}, implement a form of approximate dynamic programming without requiring explicit knowledge of every environment state transition.
Instead, they only require access to a simulation of the environment and learn through trial-and-error and continuous feedback loops, making them compatible with a larger range of modeling assumptions.
Moreover, RL is compatible with modeling policies using deep neural networks, which can model a large class of functions and asymptotically are universal function approximators.
Deep neural network policies can learn nonlinear predictive patterns over high-dimensional state-action spaces, while deep value functions can model complex reward function landscapes over long time horizons.

Combining these features has led to significant successes. 
Notably, deep RL has achieved superhuman performance in high-dimensional, sequential games \citep{silver2017mastering}, including in multi-agent settings \cite{vinyals2019grandmaster,OpenAI_dota}. 
These results suggest deep MARL can learn (approximate) strategic equilibria in complex settings. 
Pertinent to our setting, we do not have to approximate non-linear \DGEmodel{} dynamics, as deep RL has been successful in highly nonlinear environments~\citep{tassa2018deepmind}.

\paragraph{Economic Analysis using RL.}
A small but growing number of works have explored the use of RL for economic analysis, although \emph{deep RL is still not commonly used in economics}.
To our knowledge, our work is the first application of deep \emph{multi-agent} RL to \DGEmodel{} models where all agents learn.

The AI Economist used two-level RL to design optimal taxes to improve social welfare in spatiotemporal simulations, where both agents and governments use RL policies~\citep{zheng2020ai}. 
Two-level RL also yields interpretable economic and public health policies in pandemic simulations \citep{trott2021building}.
\citet{danassis2021achieving} studies harvesters who work in a common fishery and a centralized price setter, and demonstrate that an RL price setter can outperform prices found through market equilibrium on a number of metrics, including social welfare, fairness, and sustainability. 
\citet{radovic2021revealing} simulate oil company investments while transitioning away from hydrocarbons; they find that good policies for oil companies involve rapidly investing in renewable energy.

Other contemporary work includes \citet{chen2021deep}, which studies monetary policy with a single representative household RL agent,
and \citet{hill2021solving}, which learns the value function of consumers in \RBCmodel{} models.
These works are more limited, compared to ours.
First, they use RL for one agent type only, while the other agents (e.g., firms) use simple and fixed policies.
Second, they assume markets always clear at each time step, i.e., prices are manually set to ensure supply and demand are balanced.
However, this is an unrealistic assumption and causes slow simulations, requiring solving a nonlinear optimization problem at each timestep.

Perhaps most similar to our work is \citet{sinitskaya2015macroeconomies}, which studies an economy with homogeneous consumer-workers and price- and wage-setting firms (but no government). 
Here, a few consumers and firms learn using tabular Q-learning. 
Even in this small-scale setting, they observe that learning dynamics can collapse to trivial solutions with no production or consumption, similarly to our work.
Furthermore, they partially enforce market clearing by matching supply and demand for labor and goods in a double auction, but do not enforce constraints on labor availability requirements to produce demanded goods. 
In contrast, we do not enforce market clearing constraints at all, simply rationing goods if there are not enough. 
Moreover, in our \DGEmodel{} model, consumers and firms are heterogeneous, e.g., firms have different production functions and produce distinct goods.

\paragraph{Finding Equilibria in Games using Machine Learning.}
Several streams of work have explored using ML techniques to find equilibria in games. 
\citet{leibo_multi-agent_2017} studied meta-game equilbria in sequential social dilemmas, showing that RL policies can be segmented into those who effectively cooperate or defect.
In the domain of imperfect-information games, counterfactual regret minimization has yielded superhuman poker bots \citep{brown2018superhuman}. 
Empirical game-theoretic analysis studies equilibria through agent simulations \citep{wellman2006methods}, but is limited to games with interchangable agents with identical affordances \citep{tuyls2018generalised} and does not easily scale to settings with heterogeneous agents.

In extensions of deep RL, higher-order gradients \citep{foerster_learning_2017} and first-order gradient adjustments \citep{balduzzi_mechanics_2018} have been studied to promote convergence to non-trivial equilibria.
However, these methods make strong assumptions on what agents know, e.g., that agents can see the policy weights of other agents and/or know the full reward function, and may even change the equilibria of the game.

\subsection{Theoretical Analysis}
\paragraph{Different Equilibrium Types.}
\DGEmodel{} models may support many forms of equilibria.
First, assuming all agents act simultaneously, one may analyze (repeated) Nash equilibria, where no rational agent is incentivized to deviate unilaterally. 
There may also be asymmetric Stackelberg-like equilibria, where a leader (e.g., the government) acts first (e.g., sets taxes), and the followers (e.g., consumers and firms) respond \cite{zheng2020ai}.
Generally, finding all equilibria of any type in general-sum, sequential, imperfect-information games with many agents is an open challenge \citep{bai2021sampleefficient}. 
For instance, with 2 or more (heterogeneous) followers, finding the Stackelberg best-response to a fixed leader requires finding multi-agent equilibria for the followers. 
This is computationally expensive and there is no known provably convergent algorithm.

Instead, we use MARL to converge to a stable solution and analyze best-responses to evaluate to what extent it is an equilibrium.
Several works have studied the use of RL techniques in such settings \citep{wang2019deep,trejo2016adapting,kamra2018policy,bai2021sampleefficient}, although do not consider the complexity of \DGEmodel{} models.

\paragraph{Theoretical Guarantees for Convergence and Bounds on Approximate Equilibria.} 
Guarantees of convergence for model-free policy optimization are difficult to come by. 
Some theorems only apply to tabular policies (as in \citet{srinivasan2018actor}, and where convergence to Nash is not even guaranteed due to a non-sublinear regret bound). 
Other cases also apply only to very limited families of games, as in \citet{zhang2019policy}, which deals only with zero-sum linear-quadratic games. 
We study general-sum games with a large number of agents with neural network behavioral policies.
All three of these properties mean that convergence guarantees to an equilibrium are beyond the reach of theory.
To test whether a learned strategy profile is at an $\epsilon$-Nash-equilibrium, one can attempt to compute a best-response for each agent and measure $\epsilon$.
However, the non-convexity and dimensionality of neural network loss landscapes means that there are no perfect best-response oracles in practice.
In our work, we use single-agent RL and gradient-based optimization to see whether individual agents can improve their utility unilaterally and hence measure $\epsilon$.
Although the found solutions and best-responses align with economic intuitions, it is beyond the scope of this work to establish theoretical guarantees on RL as a best-response oracle, e.g., deriving upper bounds on $\epsilon$.
\begin{figure}
\centering
\includegraphics[width=\linewidth]{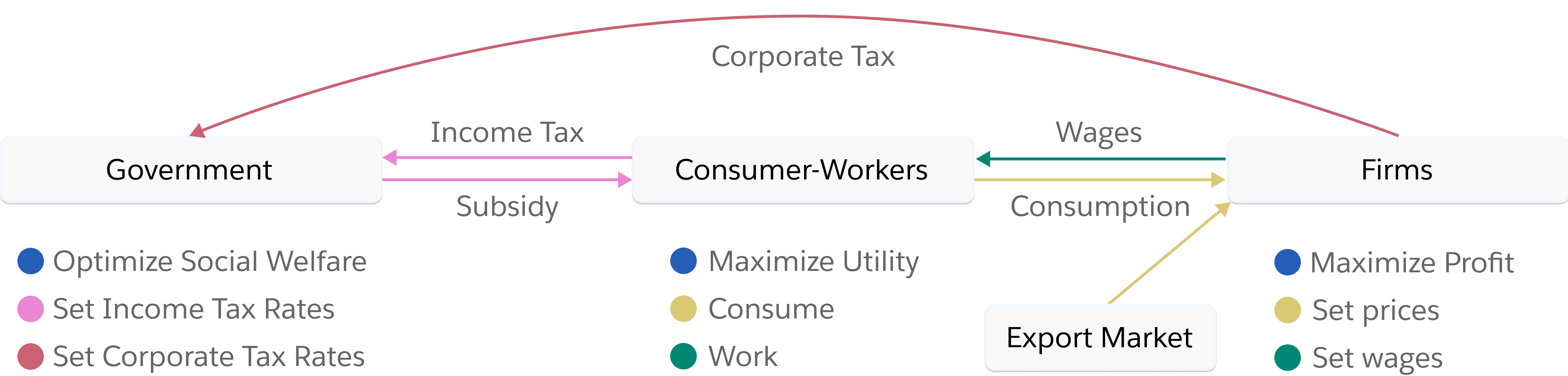}
\caption{
\textbf{\RBCmodel{} model with consumers, firms, and governments.}
Arrows represent money flow.
Consumer-workers earn wages through work and consume goods from firms. 
They also strategically choose \emph{which} firm to work for and \emph{which basket of goods} to buy, but this is not explicitly visualized.
Firms produce goods, pay wages, and set a price for their goods. 
They also invest a fixed fraction of profits to increase capital.
The government taxes labor income and firm profits, and redistribute the tax revenue through subsidies to the consumer-workers.
Firms can also sell goods to an external export market, which acts as a price-taker that is willing to consume goods at any price.
}
\label{Figure:DGE-overview}
\end{figure}

\section{
    Real-Business-Cycle Models
}\label{Section:Framework}
\RBCmodel{}s are a representative \DGEmodel{} model in which consumers earn income from labor and buy goods, firms produce goods using capital and labor, and the government taxes income and profits \citep{PIERREDANTHINE19931}, see Figure \ref{Figure:DGE-overview}.
\RBCmodel{} models are stylized and may not fully describe reality \citep{summers1986some}. 
However, \RBCmodel{} models are a suitable environment to validate the use of RL, as they feature heterogenous agents with nonlinear economic interactions, making it challenging to find equilibria.
Below, we describe the \RBCmodel{} dynamics.

At a high level, our model includes worker-consumers, price-and-wage setting firms, and a government who sets tax rates and redistributes. A key point about our model is that we do not assume as part of the environment that prices and wages are set so that markets clear at each time step -- that is, goods may be overdemanded and firms are free to set prices and wages lower or higher than would balance supply and demand. These assumptions are an essential part of the techniques used to derive analytic solutions -- avoiding this modeling choice requires using other techniques.

\paragraph{Agent Types.}
Formally, our \RBCmodel{} model can be seen as a Markov Game (MG) \citep{littman_markov_1994} with multiple agent types and partial observability.
An MG has finite-length episodes with $\eplen$ timesteps, where each timestep represents a quarter of 3 months.
At each timestep $\xtimestept$, we simulate
\begin{itemize}
\item firms and their goods who use the labor of the worker-consumers to each produce a different good, indexed by $\idxi \in I$,
\item consumers, indexed by $\idxj \in J$, who work and consume goods, and
\item a government who sets a tax rate on income from labor and on revenue from selling goods.
\end{itemize}
Each agent iteratively receives an observation $\ob_{\idxi, \xtimestept}$ of the world state $\st_{\xtimestept}$, executes an action $\ac_{\idxi, \xtimestept}$ sampled from its policy $\pol_{\idxi}$, and receives a reward $\rew_{\idxi, \xtimestept}$.
The environment \emph{state} $\St$ consists formally of all agent states and a general world state.
At each timestep $\xtimestept$, all agent simultaneously execute actions.
However, some actions only apply to the next timestep $\xtimestept + 1$.
For example, the government sets taxes \emph{that will apply at the next timestep $\xtimestept + 1$}.
These tax rates are observable by the firms and consumer-workers at timestep $\xtimestept$; hence, they can condition their behavioral policy in response to the government policy.
Similarly, at timestep $\xtimestept$, firms set prices and wages that will be part of the global state at the next $\xtimestept + 1$.
This setup is akin to, though not exactly corresponding to, a Stackelberg leader-follower structure, where, for example, the government (leader) moves first, and the firms and consumer (followers) and move second \citep{von2010market}.
This typically gives a strategic advantage to the followers: they have more information to condition their policy on.

\paragraph{Consumer-Workers.}
Individual people both consume and work; we will refer to them as \emph{consumer\nobreakdash-workers}.
At each timestep $\xtimestept$, person $\idxj$ works $\xhoursn_{\idxj, \xtimestept}$ hours and consumes $\xconsumptionc_{\idxi, \idxj, \xtimestept}$ units of good $\idxi$.
Each person chooses to work for a specific firm $\idxi$ at each timestep.
Consumer-workers also (attempt to) consume $\xtryconsumptionc_{\idxi, \idxj, \xtimestept}$.
Each firm's good has a price $p_i$ (set by the firms, described below), and the government also sets an income tax rate.
However, consumers cannot borrow or outspend their budget: if the cost of attempted consumption exceeds the budget, then we scale consumption so that $\sum_{i} \xpricep_i \xtryconsumptionc_{t,i,j} = \xbudgetb_{j}$.

Moreover, the realized consumption depends on the available inventory of goods ($y_{i,t}$, described below).
The total demand for good $\idxi$ is $\xtryconsumptionc_{\idxi, \xtimestept} = \sum_\idxj \xtryconsumptionc_{\idxi, \idxj, \xtimestept}$.
If there is not enough supply, we ration goods proportionally: 
\eq{
\xconsumptionc_{\idxi, \idxj, \xtimestept} = \min \brck{1,
\frac{\xinventoryS_{\idxi, \xtimestept}}{\xtryconsumptionc_{\idxi, \xtimestept}}
} \xtryconsumptionc_{\idxi,\idxj, \xtimestept}.
}
Consuming and working change a consumer's budget $\xbudgetb_{\idxj, \xtimestept}$.
Consumer $\idxj$ has labor income $\xincomeI_{\idxj, \xtimestept} = \sum_i \xhoursn_{\idxi, \idxj, \xtimestept} \xwagew_{\idxi, \xtimestept}$;
each firm pays a wage $\xwagew_{\idxi, \xtimestept}$.
The cost of consumption is $\sum_i \xpricep_{\idxi, \xtimestept} \cdot \xconsumptionc_{\idxi, \idxj, \xtimestept}$
Moreover, with tax rate $\tau_t$, workers pay income tax $\incometaxrate_{\xtimestept} \cdot \xincomeI_{\idxj, \xtimestept}$; the total tax revenue $\xtaxrevR_\xtimestept$ (which also includes taxes on the firms, described below) is redistributed evenly back to workers. 
In all, consumer budgets change as: 
\eq{
\xbudgetb_{\xtimestept + 1, \idxj} = \xbudgetb_{\idxj, \xtimestept} + (1 - \incometaxrate_{\xtimestept}) \xincomeI_{\idxj, \xtimestept} + \frac{\xtaxrevR_\xtimestept}{\NumConsumers} - \sum_i \xpricep_{\idxi, \xtimestept} \cdot \xconsumptionc_{\idxi, \idxj, \xtimestept}.
}
Each consumer optimizes its behavioral policy to maximize utility:
\eq{
    &\max_{\policy_{\idxj}} \mathbb{E}_{\xconsumptionc, \xhoursn \sim \policy_{\idxj}}\brcksq{
        \sum_{\xtimestept}\df_\consumersymbol^{\xtimestept}\sum_{\idxi} \util(\xconsumptionc_{\idxi, \idxj, \xtimestept}, \xhoursn_{\idxi, \idxj, \xtimestept},  \laborcoeff_\idxj)
    }, \quad%
    \util(\xconsumptionc, \xhoursn, \laborcoeff) = \frac{\brck{\xconsumptionc + 1}^{1-\eta} - 1}{1 - \eta} - \frac{\laborcoeff}{2} \xhoursn_\xtimestept,
}
where $\df_\consumersymbol$ is the consumer's discount factor.
The utility function is a sum of isoelastic utility over consumption and a linear disutility of work with coefficient $\laborcoeff_\idxj$ that can vary between workers.

\paragraph{Firms.}
At each timestep $\xtimestept$, a firm receives labor from workers, produces goods, sells goods, and may invest in capital. 
At each timestep $\xtimestept$, it sets a price $\xpricep_{t+1, \idxi}$ for its good and chooses a wage $\xwagew_{t+1, \idxi}$ to pay, both \emph{effective at the next timestep $\xtimestept + 1$}.
If a firm invests $\xcapinvestk_{\idxi, \xtimestept}$ in capital, its capital increases as $\xcapitalk_{\idxi, \xtimestept+1} = \xcapitalk_{\idxi, \xtimestept} + \xcapinvestk_{\idxi, \xtimestept}$.
Using its capital $\xcapitalk_{\idxi, \xtimestept}$ and total labor $\xtothoursN_{\idxi, \xtimestept} = \sum_\idxj \xhoursn_{\idxi, \idxj, \xtimestept}$ (hours worked), a firm $\idxi$ produces $\xproductionY$ units of good $\idxi$, modeled using the \emph{production function}:
\eq{
\xproductionY_{\idxi, t} = \prodfuncfactor_i \xcapitalk_{\idxi, \xtimestept}^{1-\prodfuncalpha} \xtothoursN_{\idxi, \xtimestept}^\prodfuncalpha,
}
where $0 \leq \prodfuncalpha \leq 1$ sets the importance of capital relative to labor.
Also, consumers buy $\xtotconsumptionC_{\idxi, \xtimestept} = \sum_\idxj \xconsumptionc_{\idxi, \idxj, \xtimestept}$ units of good $\idxi$.
Accordingly, inventories change as $\xinventoryS_{t+1, \idxi} = \xinventoryS_{\idxi, \xtimestept} + \xproductionY_{\idxi, \xtimestept} - \xtotconsumptionC_{\idxi, \xtimestept}$.
Inventories are always positive, as only actually produced goods can be consumed.
The firms receive a profit (or loss) $\xprofitP$, pay taxes on their profit, and experience a change in their budget $\xbudgetb$: 
\eq{\label{Eq:Firm Profit}
\xprofitP_{\idxi, \xtimestept} = \xpricep_{\idxi, \xtimestept} \xtotconsumptionC_{\idxi, \xtimestept} - \xwagew_{\idxi, \xtimestept} \xtothoursN_{\idxi, \xtimestept} - \xcapinvestk_{\idxi, \xtimestept},\, \xbudgetb_{t+1, \idxi} = \xbudgetb_{\idxi, \xtimestept} + (1 - \corptaxrate_{\xtimestept})\xprofitP_{\idxi, \xtimestept},
}
where $\corptaxrate$ is the corporate tax rate. The government receives $\corptaxrate_{\xtimestept} \xprofitP_{\idxi, \xtimestept}$. 
Firms may borrow and temporarily be in debt (negative budget), but should have non-negative budget at the end of an episode (\emph{no-Ponzi condition}).
This may allow firms to invest more, which may lead to higher future economic growth.
Each firm optimizes its behavioral policy to maximize profits as in Equation \ref{Eq:Firm Profit}:
\eq{
    &\max_{\policy_{\idxi}} \mathbb{E}_{\xpricep, \xwagew, \xcapinvestk \sim \policy_{\idxi}}\brcksq{
        \sum_{\xtimestept}\df_\firmsymbol^{\xtimestept} \xprofitP\brck{\xpricep_{\idxj, \xtimestept}, \xwagew_{\idxj, \xtimestept}, \xcapinvestk_{\idxj, \xtimestept}}
    },
}
where $\df_\firmsymbol$ is the firm's discount factor.
The firm gets a negative penalty if it violates \emph{no-Ponzi}.

\paragraph{Government.}
The government, or social planner, indexed by $\idxplanner$, sets corporate and income tax rates, and receives total tax revenue $\xtaxrevR_{\xtimestept} = \corptaxrate_{\xtimestept} \sum_\idxj\xincomeI_{\idxj, \xtimestept} + \incometaxrate_{\xtimestept} \sum_\idxi\xprofitP_{\idxi, \xtimestept}$. 
As a modeling choice, this revenue is redistributed evenly across consumer-workers; as such, the government's budget is always 0.
The government optimizes its policy $\policy_{\idxplanner}$ to maximize social welfare:
\eq{
    \max_{\policy_{\idxplanner}} \mathbb{E}_{\incometaxrate, \corptaxrate \sim \policy_{\idxplanner}}\brcksq{
        \sum_{\xtimestept}\df_\governmentsymbol^{\xtimestept} \SWF\brck{\incometaxrate_{\xtimestept}, \corptaxrate_{\xtimestept}, \St_{\xtimestept}}
    },
}
where $\SWF\brck{\St_{\xtimestept}}$ is the social welfare at timestep $\xtimestept$, and $\df_\governmentsymbol$ is the government's discount factor. 
In this work, we consider two definitions of social welfare (although many other definitions are possible): 
1) \emph{total consumer utility} (after redistribution), or
2) \emph{total consumer utility and total firm profit}.

\paragraph{Open and Closed Economies via Export Markets.}
We consider both open and closed economies. 
In the open economy, firms can also sell goods to an external market which acts as a \emph{price-taker}: their demand does not depend on the price of a good.
Operationally, export happens after worker-consumers consume. 
The export market has a minimum price $\xpricep_{\text{export}}$ and a cap $\xquantityq_{\text{export}}$.
If the price of good $\idxi$ is greater than the minimum price ($\xpricep_{\idxi, \xtimestept} > \xpricep_{\text{export}}$) then the additional export consumption is
$\xconsumptionc_{\xtimestept, \text{export}} = \min(\xquantityq_{\text{export}}, \xinventoryS_{\idxi, \xtimestept} - \xtotconsumptionC_{\idxi, \xtimestept})$, at price $\xpricep_{\idxi, \xtimestept}$, i.e., the exported quantity is insensitive to the price.

From a learning perspective, the export market prevents firms from seeing extremely low total demand for their good, e.g., when prices are exorbitantly high and consumers do not want or cannot consume the good.
In such cases, an on-policy learner that represents a firm may get stuck in a suboptimal solution with extremely high prices and no production as consumers cease to consume in response.

\section{
Key Simulation Implementation Details
}
\label{Section:Implementation Details}
In general, experimental outcomes can depend significantly on the implementation details; we outline several key implementation details hereafter.
In addition, all simulation and training settings can be found in Table \ref{Table:Simulation Parameters} and Table \ref{Table:Training Hyperparameters}.

\paragraph{Budget Constraints.}
We implement budget constraints on the consumers by proportionally scaling down the resulting consumption of all goods to fit within a consumer's budget.
Thus, the consumer actions really represent attempted consumption -- if the budget is small or stock is limited, the actual consumption enjoyed by the consumer may be lower.
Firm budgets are allowed to go negative (borrowing money).
However, because the firm's goal is to maximize profit, they are incentivized to take actions will will be profitable, increasing their budget.

\paragraph{Scaling of Observables.}
The scales of rewards and state variables can vary widely in our simulation, even within time steps in a single episode. 
If the scales of loss functions or input features are very large or small, learning becomes difficult. 
We directly scale rewards and some state features by constant factors. For certain state features which have very large ranges (item stocks and budgets) we encode each digit of the input as a separate dimension of the state vector.

\paragraph{GPU Implementation}
We followed the WarpDrive framework \citep{lan2021warpdrive} to simulate the \DGEmodel{} model and run MARL on a single GPU.
We implemented the \DGEmodel{} dynamics as a CUDA kernel to leverage the parallelization capabilities of the GPU and increase the speed at which we can collect samples. 
We assigned one thread per agent (consumer, firm, or government); the threads communicate and share data using block-level shared memory.
Multiple environment replicas run in multiple blocks, allowing us to reduce variance by training on large mini-batches of rollout data.
We use PyCUDA \citep{klockner2012pycuda} to manage CUDA memory and compile kernels.
The policy network weights and rollout data (states, actions, and rewards) are stored in PyTorch tensors; the CUDA kernel reads and modifies these tensors using pointers to the GPU memory, thereby working with a single source of data and avoiding slow data copying.

\begin{table}[t]
    \centering
    \small
    \begin{tabular}{lcl}
         \textbf{Parameter} & \textbf{Symbol} & \textbf{Values}  \\
         Labor disutility  & $\laborcoeff$  &  0.01 \\
         Pareto quantile function scale parameter & - & 4.0 \\
         Initial firm endowment & $\xbudgetb$ & 2200000 \\
         Export market minimum price & - & 500 \\
         Export market maximum quantity & - & 100 \\
         Production function values & $\prodfuncalpha$ & $0.2, 0.4, 0.6, 0.8$ \\
         Initial capital & $\xtotcapitalK$ & 5000 or 10000 \\
         First round wages & $\xwagew$ & 0 \\
         First round prices & $\xpricep$ & 1000 \\
         Initial inventory & $\xinventoryS$ & 0 \\
    \end{tabular}
    \caption{\textbf{Simulation Parameters.} }
    \label{Table:Simulation Parameters}
\end{table}
\begin{table}[t]
    \centering
    \small
    \begin{tabular}{ll}
         \textbf{Parameter} &  \textbf{Values}  \\
         Learning Rate   &  0.001 \\
         Learning Rate (Government) & 0.0005 \\
         Optimizer & Adam \\
         Initial entropy & 0.5 \\
         Minimum entropy annealing coefficient & 0.1 \\
         Entropy annealing decay rate & 10000 \\
        Batch Size & 128 \\
        Max gradient norm & 2.0 \\
        PPO clipping parameter & 0.1 or 0.2 \\
        PPO updates & 2 or 4 \\
        Consumer reward scaling factor & 5 \\
        Firm reward scaling factor & 30000 \\
        Government reward scaling factor & 1000
    \end{tabular}
    \caption{\textbf{Training Hyperparameters.} }
    \label{Table:Training Hyperparameters}
\end{table}

\paragraph{Implementation Details.} 
Furthermore, we outline several key implementation details.
\begin{itemize}
    \item For consumers, consumption choices range from 0 to 10 units for each good and work choices from 0 to 1040 hours in increments of 260.
    \item Consumers have a CRRA utility function with parameter 0.1, and a disutility of work of 0.01.
    \item For firms, price choices range from 0 to 2500 in units of 500; wage choices from 0 to 44 in units of 11.
    \item The 10 firms are split into two groups, receiving either 5000 or 10000 units of capital. Within these groups, firms receive a production exponent ranging from 0.2 to 0.8 in increments of 0.2. Thus each firm has a different production ``technology''.
    \item Firms invest 10\% of their available budget (if positive) in each round to increase their capital stock.
    \item Government taxation choices range from 0 to 100\% in units of 20\%, for both income tax and corporate tax rates.
    \item The government can either value only consumers when calculating its welfare (``consumer-only'') or value welfare of both consumers and firms (``total''), with firm welfare downweighted by a factor of 0.0025 (to be commensurate with consumers).
    \item We set the minimum price at which firms are willing to export to be either 500 or 1000, and the quota for each firm's good to a variety of values: 10, 50, 100, or 1000. 
    \item For consumers, consumption choices range from 0 to 10 units for each good and work choices from 0 to 1040 hours in increments of 260.
\end{itemize}

\paragraph{Agent Observations.}
The environment \emph{state} $\St$ consists formally of all agent states and a general world state.
Each agent observes can observe their own information and the global state:
\eq{
\St_\textrm{global} = \brck{
\xtimestept,
\brckcur{\xinventoryS_{\idxi, \xtimestept}}_\idxi, 
\brckcur{\xpricep_{\idxi, \xtimestept}}_\idxi, 
\brckcur{\xwagew_{\idxi, \xtimestept}}_\idxi, 
\brckcur{\xobso_{\idxi, \xtimestept}}_\idxi
}.
}
Here $\xinventoryS_{\idxi, \xtimestept}$ is the available supply of good $\idxi$, $\xpricep_{\idxi, \xtimestept}$ is the price, $\xwagew_{\idxi, \xtimestept}$ is the wage. The extra information $\xobso_{\idxi, \xtimestept}$ includes whether good $\idxi$ was overdemanded at the previous timestep and tax information.

In addition, consumer-workers observe private information about their own state: 
$\brck{ 
\xbudgetb_{\idxi, \xtimestept}, 
\laborcoeff
}$
A firm $\idxi$ also observes its private information:
$\brck{ 
\xbudgetb_{\idxi, \xtimestept}, 
\xcapitalk_{\idxi, \xtimestept}, 
\brck{0, \ldots, 1, \ldots, 0}, 
\prodfuncalpha
}$, 
including a one-hot vector encoding its identity and its production function shape parameter $\prodfuncalpha$.
The government only sees the global state.

\section{Reinforcement Learning Algorithm for a Single Agent}
Firms and governments use 3-layer fully-connected neural network policies $\policy\brck{\V{\ac}|\V{\st}}$, each layer using 128-dim features, that map states to action distributions. 
Consumer policies are similar, using separate heads for each action type, i.e., the joint action distribution is factorized and depends on a shared neural network feature $\nnfeat_\xtimestept(\st_\xtimestept)$: $\policy(\ac_1, \ac_2, \ldots | \st) = \policy(\ac_1 | \nnfeat, \st) \policy(\ac_2 | \nnfeat, \st)\ldots$ (omitting $\xtimestept$ and $\st_\xtimestept$ for clarity).
Any correlation between actions is modeled implicitly through $\nnfeat_\xtimestept$.

There is a single policy network for each agent type, shared across the many agents of that type. To distinguish between agents when selecting actions, agent-specific features (parameters like the disutility of work, production parameters, and for firms, simply a one-hot representation of the firm) are included as part of the policy input state. Thus, despite a shared policy for each agent type, we model some degree of heterogeneity among agents.
We also learn a value function $\Val{}\brck{\nnfeat_\xtimestept}$ for variance reduction purposes.
We compare policies trained using policy gradients \citep{Williams:1992:SSG:139611.139614} or PPO \cite{schulman2017proximal}.

\paragraph{RL parameter updates} 

We now describe the RL parameter updates for any given agent type.

Given a sampled trajectory of states, actions, and rewards $s_t, a_t, r_t$ for $t$ from 0 to the end time step $T$, we have the empirical return $G_t = \sum_{k=0}^{T - t} \gamma^k r_{t+k+1}$, the total discounted future rewards from time step $t$.
The goal of the value function network is to accurately predict $G_t$ from a given state $s_t$. 
We use a Huber loss function $\ell$ to fit the value network, $\ell(\Val_\beta{(s_t)} - G_t)$, and the value weights are updated as $\beta_{t+1} = \beta_t - \eta \nabla_\beta \sum_{t=0}^T \ell(\Val_\beta{(s_t)} - G_t)$, where $\eta$ is the step size for gradient descent.
Given the value function network's predictions, we can then define the \textit{advantages} $A_t = G_t - \Val(s_t)$ and their centered and standardized versions $\hat{A}_t = \brck{A_t - \mathbb{E}_\pi\brcksq{A}} / \text{std}(A)$.

For the policy gradient approach, the optimization objective for the policy is:
\eq{
    \max_\theta \mathbb{E}_{\pi_\theta}\brcksq{A_t} + \alpha H(\pi_\theta),
}
where $H(\pi) = \mathbb{E}_\pi\brcksq{-\log\pi}$ is the entropy of $\pi$, $\alpha$ is the weight for entropy regularization (which may be annealed over time), and $\pi_\theta$ is the policy with parameters $\theta$.
The true policy gradient for the first term is $\mathbb{E}_{\pi_\theta}\brcksq{A_t\nabla_\theta\log\pi_\theta}$, which we estimate using sampled trajectories.
For a single sampled trajectory, the full policy weight update is:
\eq{
\theta_{t+1} = \theta_t - \nabla_\theta \left(\sum_{t=0}^T \hat{A}_t \log \pi_\theta(a_t | s_t) + \alpha H(\pi_\theta(\cdot | s_t))\right),
}
where $H(\pi(\cdot | s_t))$ is the entropy of the action distribution at a particular state.
In practice, we sample multiple trajectories and compute a mini-batch mean estimate of the true policy gradient.
In addition, we use proximal policy optimization (PPO), a more stable version of the policy gradient which uses a surrogate importance-weighted advantage function $A_{PPO}$ in the policy objective:
\eq{
   A_{PPO} = \min\left( \frac{\pi_\theta(a_t|s_t)}{\pi_{\text{old}}(a_t | s_t)} \hat{A}_t, \text{clip}\left( \frac{\pi_\theta(a_t|s_t)}{\pi_{\text{old}}(a_t | s_t)} \right) \hat{A}_t \right),
}
which uses the current $\pi_\theta$ and policy before the last update $\pi_\textrm{old}$. 
Extreme values of the importance weights $\pi_\theta/\pi_\textrm{old}$ are clipped for stability.
Moreover, in practice, using the standardized advantages $\hat{A}_t$ and clipping gradients to bound their $\ell_2$ norm also improve stability. 
Both simulation and RL ran on a GPU using the WarpDrive approach \citep{lan2021warpdrive}, see Section \ref{Section:Implementation Details}, while our PPO implementation followed \cite{pytorchrl}.
We show pseudo-code for a single simulation and training step in Algorithm \ref{alg:trainstep}.

\begin{algorithm}[t]
\caption{A single training step at time step $t$.}
\begin{algorithmic}
\Let{$\pi_c, v_c$}{consumer policy and value network}
\Let{$\pi_f, v_f$}{firm policy and value network, prices and wages masked according to $t$}
\Let{$\pi_g, v_g$}{masked government policy and value network, tax rates masked according to $t$}

\State $\theta(t)$: disutility of work parameter, annealed over training steps

\State $w(t)$: entropy parameter, annealed over training steps according to $\max(\exp(\frac{-t}{\text{decay rate}}), 0.1)$

\Let{$s_c, a_c, r_c, s_f, a_f, r_f, s_g, a_g, r_g$}{EnvironmentSimulate($\pi_c$, $\pi_f$, $\pi_g$, $\theta(t))$}

\Let{$\pi_c, v_c$}{PPOUpdate($\pi_c, v_c, s_c, a_c, r_c, w(t)$)}
\If{$t > t_{\text{start firm}}$}
    \Let{$\pi_f, v_f$}{PPOUpdate($\pi_f, v_f, s_f, a_f, r_f, w(t - t_\text{start firm})$)}
\EndIf
\If{$t > t_{\text{start government}}$}
    \Let{$\pi_g, v_g$}{PPOUpdate($\pi_g, v_g, s_g, a_g, r_g, w(t - t_\text{start government}$)}
\EndIf

\end{algorithmic}
\label{alg:trainstep}
\end{algorithm}

\section{Stable Multi-Agent Reinforcement Learning through Structured Curricula} 
\label{Section:Training Details}

\begin{figure}[t]
    \centering
    \includegraphics[width=0.617\linewidth]{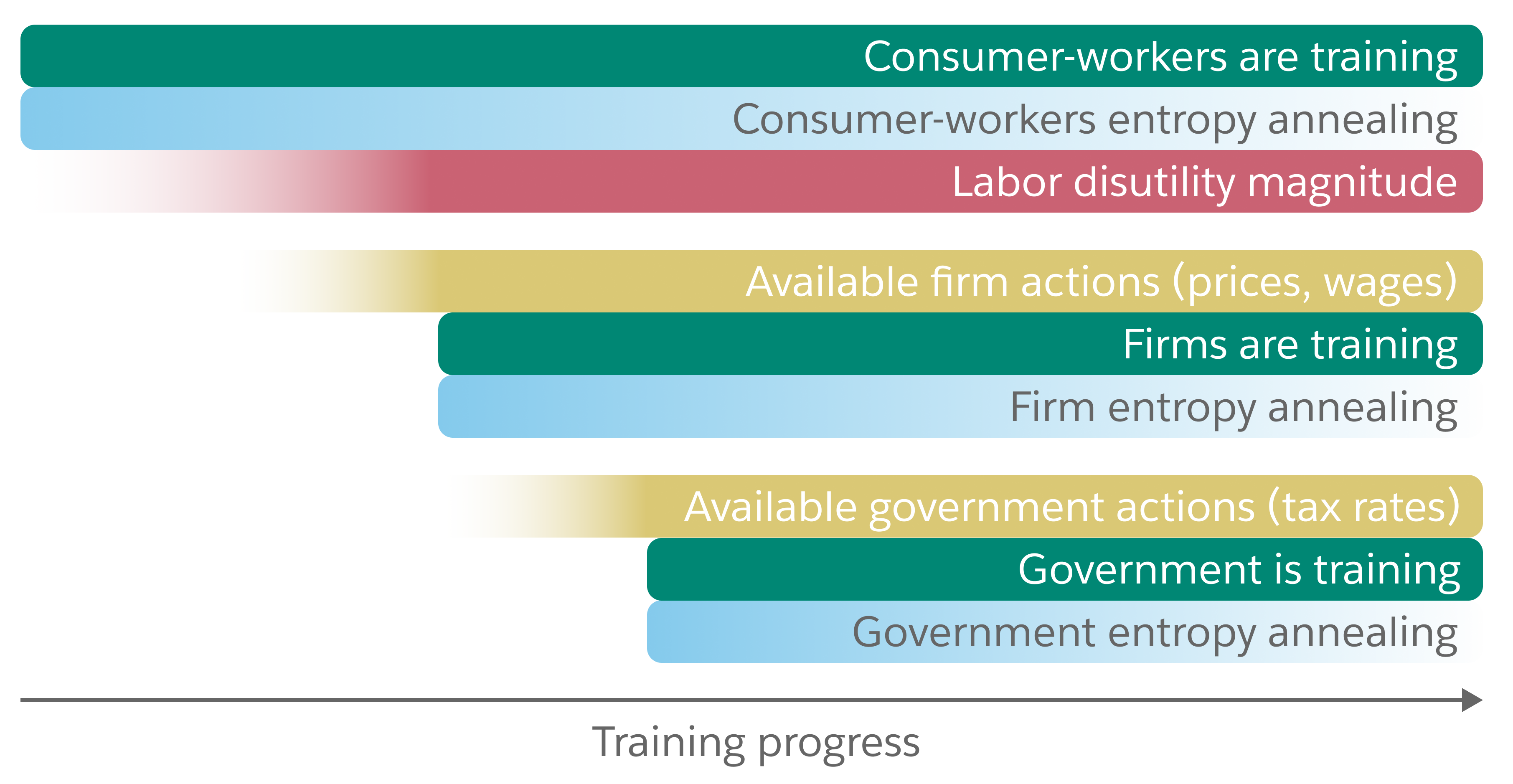}
    \caption{
        \textbf{Structured learning curricula.}
        Colored bars show how annealing activates over time.
        Consumers are always training, use a decaying entropy regularization, and increasingly experience their labor disutility. 
        Firms start with fixed prices and wages, and are slowly allowed to change these. 
        Once the full price and wage range is available, firms start to train.
        Similarly, tax rates start at zero and are slowly allowed to change. 
        Once the full tax range is available, the government starts to train.
    }
    \label{Figure:Curricula}
\end{figure}

A key idea of our approach is the use of structured multi-agent curricula to stabilize learning, where each individual agent uses standard RL.
These curricula consist of staged training, annealing of allowed actions, and annealing of penalty coefficients, see Figure \ref{Figure:Curricula}.
This method extends the approach used by \citet{zheng2020ai}.

\paragraph{Intuition for Structured Curricula.}
We define curricula based on these observations about our multi-agent learning problem:
1) during exploration, many actions may reduce utility, while few actions increase utility, 
2) high prices or tax rates can eliminate all gains in utility, even though the consumer-worker did not change its policy,
3) for stable learning, agents should not adapt their policy too quickly when experiencing large negative (changes in) utility, and
4) in a Stackelberg game, the followers (e.g., consumers, firms) should get enough time to learn their best response to the leader's policy (e.g., the government). 
We now operationalize these intuitions below.
\paragraph{Staged Learning and Action Space Annealing.}
All policies are randomly initialized.
We first allow consumers to train, without updating the policies of other agents.
Initially, firm and government actions are completely fixed;
prices and wages start at non-zero levels.
We then anneal the range of firm actions without training the randomly initialized policy.
This allows consumers to learn to best respond to the full scope of prices and wages, without firms strategically responding.
Once firm action annealing is complete, we allow the firm to train jointly with the consumers.
We then perform the same process, gradually allowing the government to increase its corporate and income tax rates, so that firms and consumers can react to a wide range of tax rates.
Once the annealing process is complete, we allow the government to train to maximize welfare.

\paragraph{Penalty Coefficient Annealing.}
In addition to the action annealing, we anneal two penalty coefficients.
First, we slowly increase the consumers' disutility of work over time, which avoids disincentivizing work early in the training process.
Second, as each agent starts training, we assign a high value (0.5) for the entropy coefficient in their policy gradient loss, and gradually anneal it down over time to a minimum of 0.1.
This ensures that when the firm or government policies start training, their ``opponent'' policies are able to learn against them without being too quickly exploited.

\paragraph{Many Local Equilibria and Convergence.} 
We expect that there are many possible solutions in our game that our approach may converge to. 
Establishing convergence is difficult in general-sum games -- the best results make use of regret-minimizing algorithms for the agents and only establish convergence to coarse-correlated equilibria~\citep{daskalakis2021near}.
In our case, convergence is hard to guarantee and may be only local since we use deep neural networks and can't ensure global optimality of our policy parameters.
We don't have theoretical bounds on the magnitude of $\epsilon$ for an $\epsilon$-Nash equilibrium, although empirically, the degree of possible improvement seems small across RL approximate best-response analysis.

Our training curricula are designed to avoid trivial solutions where the economy shuts down, but it may also introduce bias into which non-trivial solutions are reached. 
However, we observe a spectrum of qualitatively different outcomes, so we see our approach as a way to explore more solutions than are possible with simplified, e.g., linearized, models.
\begin{figure}[t]
    \centering
    \includegraphics[width=1.0\linewidth]{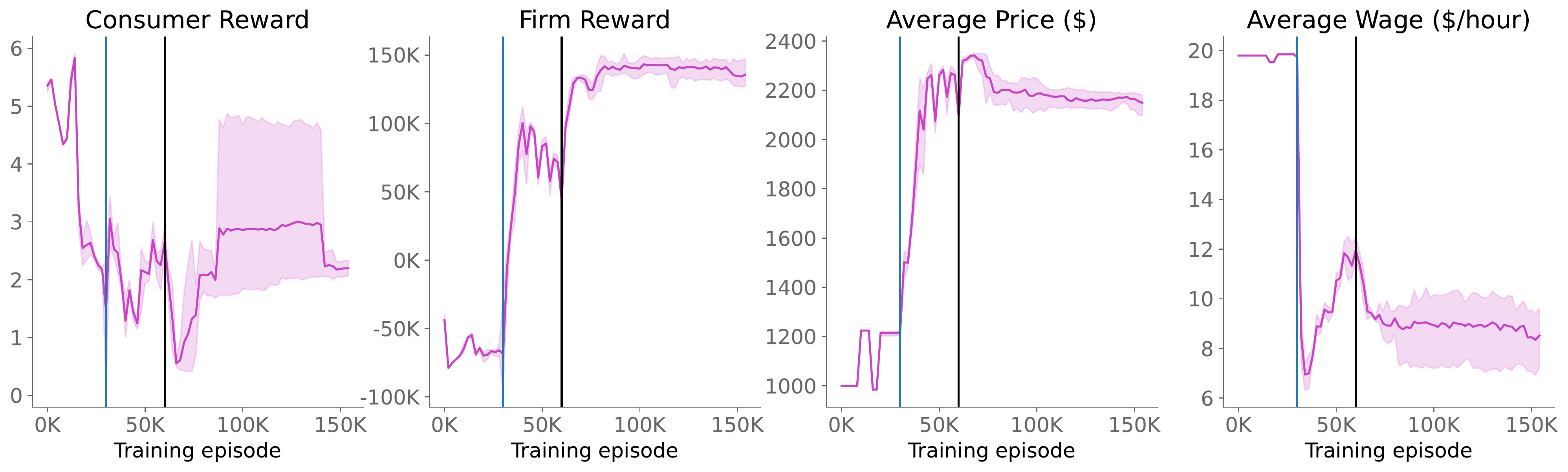}
    \caption{
        \textbf{Training progress with structured curricula in an open \RBCmodel{} model.}
        Curves show averages across 3 repetitions with different random seeds.
        In each plot, the blue (black) vertical line shows when the firms (government) start training, see Figure \ref{Figure:Curricula}. 
        \textbf{Left two plots:} Consumer and firm rewards during training.
        All runs converged to qualitatively similar outcomes.
        We've confirmed these solutions form an approximate equilibrium under an approximate best-response analysis, see Figure \ref{Figure:best-responses vs Training Time}.
        Once firms start training, their reward (profits) significantly increases. 
        When the government starts training, firms get even higher reward, as the social welfare definition includes the firms' profits. 
        \textbf{Right two plots:} Average wages and prices across firms during training.
        Firms increase prices rapidly and lower wages once they start training.
        This comes at the expense of consumer reward (utility).
        As such, this setting represents an economy in which firms have significant economic power.
    }
    \label{Figure:Rewards vs Training Time}
\end{figure}
\begin{figure}[t]
\centering
        \textbf{Approximate Best-Response Improvements}\par\medskip
    \includegraphics[ height=150pt]{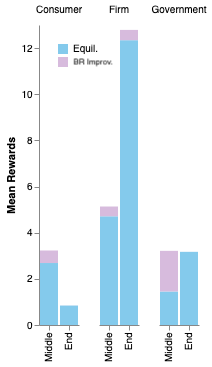}
    \includegraphics[ height=150pt]{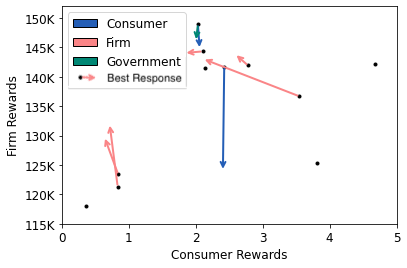}
    \centering
    \caption{
        In each plot, the agent types (consumer, firm, and government) refer to cases when only that agent type is training.
        \textbf{Left: best-response reward improvements during training.}
        The stacked bar chart shows the original mean rewards (blue) and improvement after approximate best-response training (purple).
        For firms and governments, the mean rewards are measured in units of $10^4$ and $10^3$, respectively.
        We compare the best-response improvement in the middle and at the end of training. 
        The improvement from best-response is significant in the middle and much less at the end, indicating that training is closer to an equilibrium at the end.
        \textbf{Right: outcome shifts under best-responses.} 
        We plot firm against consumer rewards on an absolute scale \emph{at training convergence} for several runs.
        We then find best-responses by further training each agent type separately. 
        Each arrow shows the shift in (consumer reward, firm reward) after best-response: blue for consumers best responding, red for firms, and green for government.
        In the figure, we display only those arrows when rewards change by more than 1\%.
        At convergence, rewards for any agent type typically do not change significantly in this best-response analysis. 
        This holds generally for the approximate equilibria reported in this work.
    }
    \label{Figure:best-responses vs Training Time}
    \label{Figure:Fixed vs RL Government}
\end{figure}
\begin{figure}
    \centering
    \includegraphics[height=150pt]{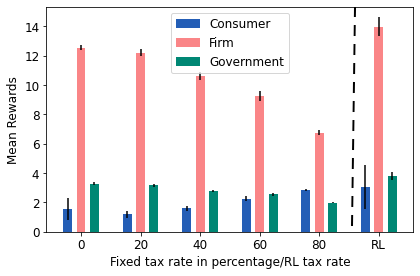}
    \caption{
        \textbf{Mean rewards under fixed and RL taxes.} 
        For firms and governments, the mean rewards are measured in units of $10^4$ and $10^3$, respectively.
        The mean rewards for the consumers increase with tax rates, whereas they decrease for the firms.
        RL taxes can improve the mean reward for both types.
        For instance, RL tax policies increase social welfare (green) by almost 15\% over the best fixed tax policy.
        }
    \label{fig:taximprovement}
\end{figure}
\begin{table}
    \centering
    \begin{tabular}{l c c c }
          & \textbf{Consumer} & \textbf{Firm} & \textbf{Government} \\
         \midrule
         Reward improvement under best-response & $< 3\%$ & $< 10\%$ & $< 1\%$ 
    \end{tabular}
    \caption{
    \textbf{Reward improvement under best-response.} 
    Best-responses are measured at the end of training as a fraction of the reward improvement during training, over 10 random seeds.
    These results represent worst-cases: due to the stochastic nature of RL, outcomes can differ significantly across individual runs.
    In fact, we observed that besides a few anomalies, overall the improvements were in fact less than 0.2\% for consumers, 5\% for firms, and 0.1\% for the government.
    }
    \label{Table:BestResponseImprovement}
\end{table}
\begin{figure}[t]
    \textbf{Learned Solutions in Closed RBC Models}\par\medskip
    \includegraphics[width=\linewidth, height=100pt]{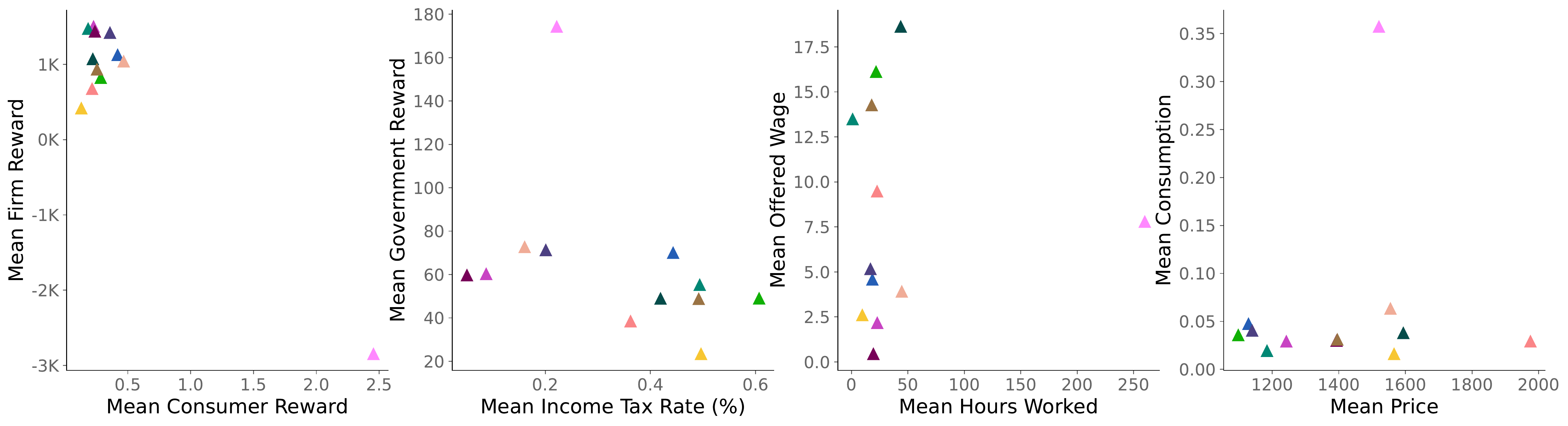}
    \centering
    \caption{
        \textbf{Outcomes at convergence for the same experiments in Figure \ref{Figure:best-responses vs Training Time}, under the closed economy.}
        Each point represents an approximate equilibrium, verified by an approximate best-response analysis.
        Points of the same color and shape correspond to the same run.
        In this closed economy, training often converges to solutions with low consumer reward and little production. 
        In particular, social welfare (government reward) does not increase with higher tax rates, average labor does not change with wages, and consumption is unchanged with price.
        An exception is a solution with significantly higher social welfare, labor, and consumption. 
        This suggests multiple equilibria do exist, but non-trivial equilibria are harder to learn in the closed economy.
    }
    \label{Figure:Learned Equilibria for Closed RBC Models}
\end{figure}

\begin{figure}
    \centering
        \textbf{Learned Solutions in Open RBC Models}\par\medskip
    \includegraphics[width=\linewidth, height=100pt]{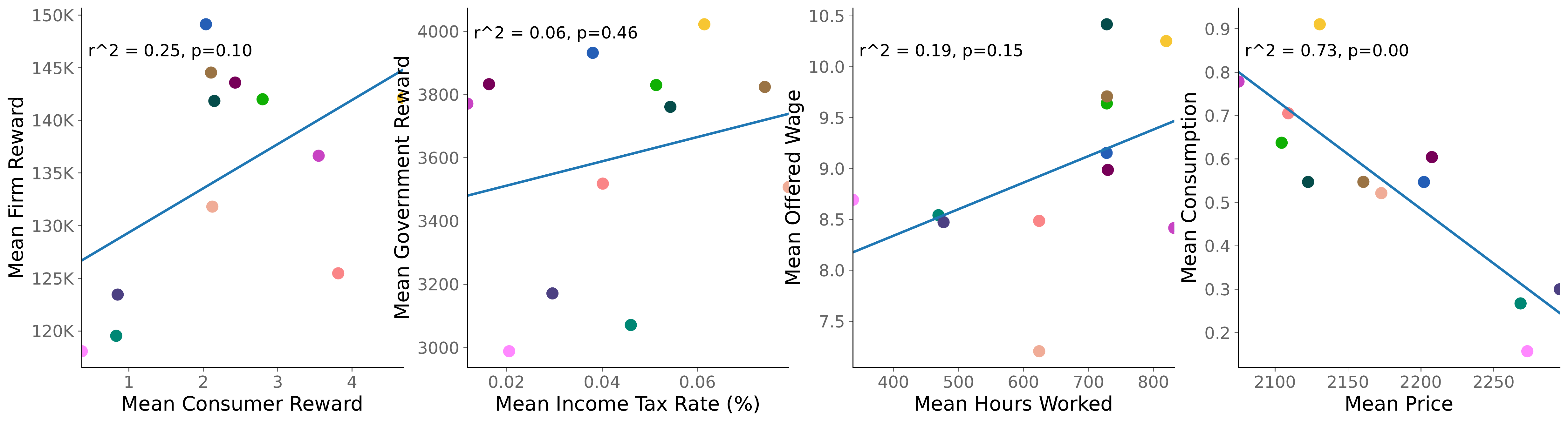}
    \caption{
        \textbf{Outcomes at convergence for the same experiments in Figure \ref{Figure:best-responses vs Training Time}, under the open economy.}
        Each point represents an approximate equilibrium, verified by an approximate best-response analysis.
        Points of the same color and shape correspond to the same run.
        We learn multiple distinct, non-trivial solutions in an open economy.
        Blue lines show linear regressions to the data.
        Consumer and firm rewards are positively correlated ($r^2 = 0.25$), e.g., if consumers earn more, they can consume more, yielding higher profits. 
        Higher prices decrease mean consumption ($r^2 = 0.73$), lower wages decrease mean hours worked ($r^2 = 0.19$), and there is no strong signal that higher taxes improve social welfare ($r^2 = 0.06$).
}
    \label{Figure:learned equilibria open}
\end{figure}
\begin{figure}[t]
    \centering
    \includegraphics[width=0.8\linewidth]{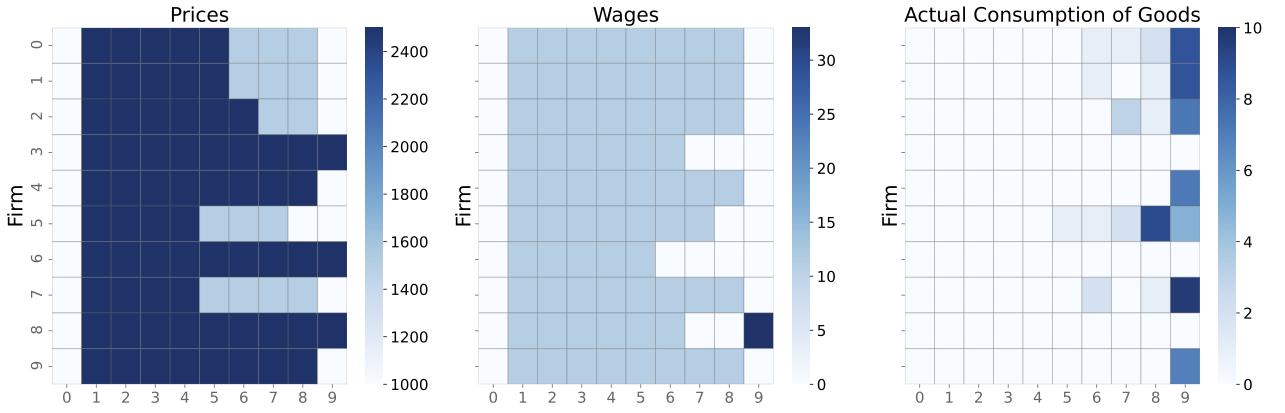}
    \includegraphics[width=0.8\linewidth]{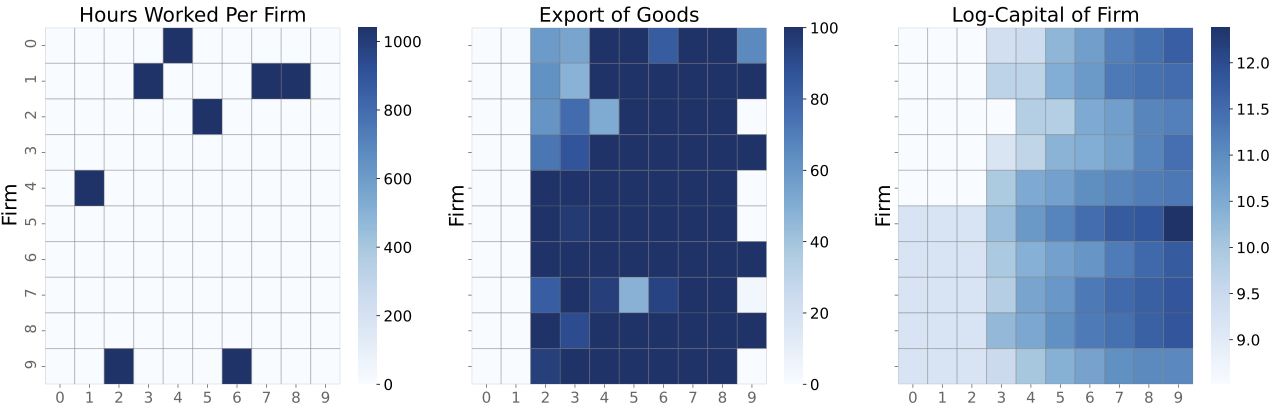}
    \includegraphics[width=.8\linewidth]{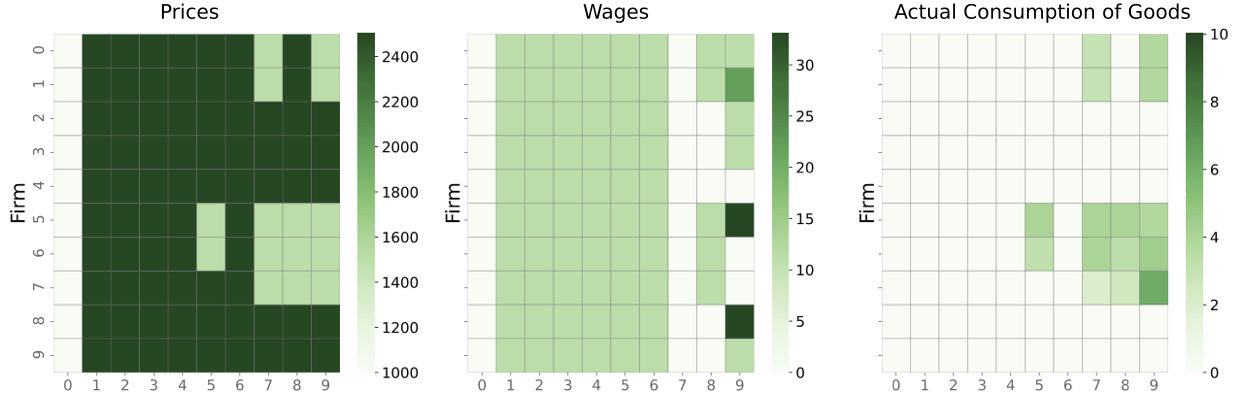}
    \includegraphics[width=.81\linewidth]{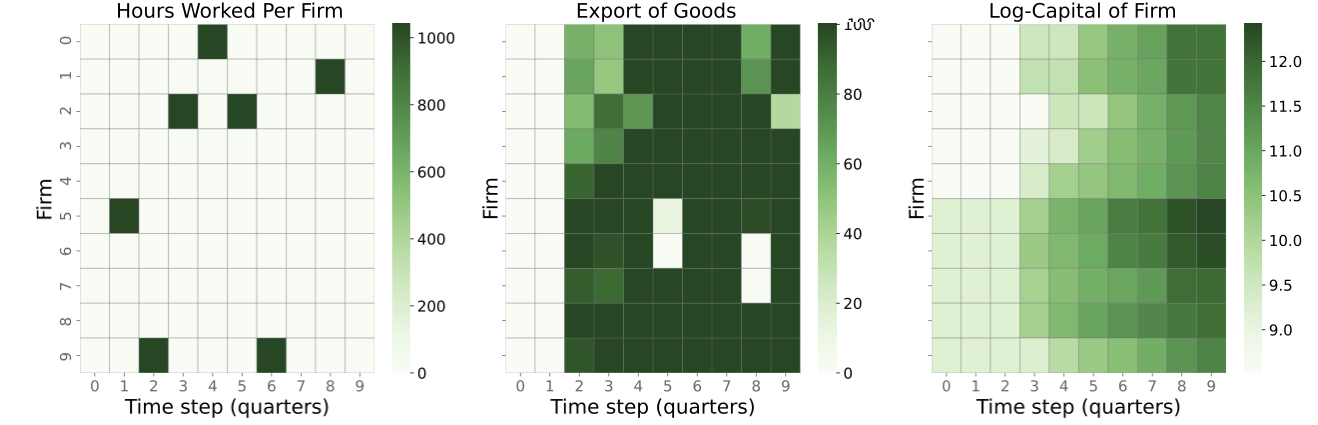}
    
    \caption{
        \textbf{Two rollouts from an open \RBCmodel{} model.}
        We show actions and states for two representative runs, in blue and green respectively.
        Despite differences in strategies, there are many qualitatively similar features.
        We observe that firms have different strategies: some set prices high and rely on exporting goods (for example, firm 3); others set prices lower and also sell to consumers (for example, firm 0).
        Consumers respond sensibly, only consuming when prices are low and mainly working when wages are not 0. 
        (Note that hours worked may be very low for some firms but are not necessarily 0.)
        The first 5 firms have a lower level of starting capital.
        Within each group, the production exponent for labor ranges from 0.2 (higher tech, more reliance on capital) to 0.8 (lower tech, more reliance on labor).
        Firms 0 and 5 have the lowest parameter of 0.2; firms 4 and 9 have the highest parameter of 0.8.
        The differences in initial firm endowments as well as evolution during the episode lead to different final capital levels.}
    \label{Figure:Episode Rollout}
\end{figure}

\section{
Analyzing Learned Strategic Behaviors in \RBCmodel{} Models
}\label{Section:Results}
We show that our approach is sound and finds spectra of meaningful solutions of \RBCmodel{} models.
We analyze to what extent these solutions are \RBCmodel{} equilibria using an approximate best-response analysis and analyze the economic relationships that emerge from the learned solutions.
We study variations of our \RBCmodel{} model with 100 consumers and 10 firms.
For all simulation parameter settings, see Table \ref{Table:Simulation Parameters}.
We repeated all experiments with 3 random seeds.

\paragraph{Structured Curricula during Training.}
Figure \ref{Figure:Rewards vs Training Time} shows a representative result of training using our structured curricula.
Although all RL agents aim to improve their objective, some agents may see rewards decrease as the system converges, e.g., the consumers in Figure \ref{Figure:Rewards vs Training Time}. 
Empirically, we found that training collapses to trivial solutions much more often without curricula.

\paragraph{Best-responses and Local Equilibrium Analysis}
We abstract the \RBCmodel{} as a normal-form meta-game between three players representing the agent types, i.e., the consumers, firms, and government.
We test whether a set of learned policies is an $\epsilon$-Nash equilibrium for the meta-game by evaluating whether or not they are approximate best-responses.
Recall that agents of the same type share policy weights, but use agent-specific inputs to their policy, hence this still models heterogeneity between agents. 

To find an approximate best-response, we train each agent type separately for a significant number of RL iterations, holding other agent types fixed, and see to what extent this improves their reward.
This measurement provides a lower bound on the $\epsilon$ for each \emph{local} equilibrium.
In general, we find empirically that best-responses improve rewards much more at the start compared to the end of training, see Figure~\ref{Figure:best-responses vs Training Time}.
The best-response method found at most small improvements in agent rewards, \emph{at the end of training}, see Table \ref{Table:BestResponseImprovement}. 
This suggests that the use of RL for a single agent type is a practical method that yields meaningful results.

For practical reasons, as we consider settings with a large number of agents (e.g., 100 consumers), we limit our analysis to evaluating \emph{meta-game} best-responses.
The meta-game is defined over agent types, and is different from the game over individual agents.
In the meta-game, a meta-agent consist of the agents of a single type and a meta-action consists of the policy prescribing actions for all agents of that type.
In particular, a meta-game best-response (where the best-response is over the choice of policy shared by \emph{all} agents of a given type) may not be an individual agent's best-response.
A meta-game best-response may feature competition-coordination dynamics between agents of the same type and may introduce both beneficial or adverse effects for some individual agents.
For instance, there may be free-riders that benefit from the collective performance while not putting in effort themselves.
On the other hand, aggregate demand between competing consumers for a limited supply of goods may mean consumers on average lose utility.
We empirically observed the latter case: overall consumer utility sometimes decreased during the best-response analysis for the consumers.

\paragraph{Comparing with Baseline Government Tax Policies.}
To show that our approach is sound, we show that RL tax policies lead to improved social welfare, compared to several manually-defined fixed tax-rate policies. 
In lieu of other potential baseline tax policy methods, which are not directly compatible with our approach due to discretization, the lack of market clearing, and other features of our model, these fixed tax policies provide a useful baseline.
Here, social welfare is defined as a weighted sum of firm and consumer rewards, as defined in Section \ref{Section:Implementation Details}. 
As observed in Figure~\ref{Figure:learned equilibria open}, RL policies generate a social welfare ranging from 3000 to 4000, depending on the solution (as noted above, multiple solutions could be reached).
Compared to that, the best social welfare achieved using fixed tax rates was 3160.
Figure~\ref{Figure:Fixed vs RL Government} shows the social welfare achieved under various fixed tax-rates, ranging from 20\% to 80\%.
We note that social welfare improves by almost 15\% for the best solution under RL tax policy over the best fixed tax-rate policy.
This shows that the RL policy can adjust taxes across different rounds to improve average social welfare.

\paragraph{Analyzing Learned Solutions}
Figure~\ref{Figure:Learned Equilibria for Closed RBC Models} visualizes and describes the learned solutions in open and closed \RBCmodel{} economies (with and without an export market).
In both cases, we find multiple qualitatively different solutions.
In particular, learning mostly converges to mostly low-welfare solutions in the closed economy, due to very low wages resulting in little labor, production, and consumption. 
We stress that we describe these observations at the level of correlations. 
The \RBCmodel{} dynamics lead to circular interactions, e.g., a change in wages can cause future changes in wages due to changing consumer and hence changing firm behavior. 
As such, disentangling cause and effect is beyond the scope of this work.

\paragraph{Relationships between Aggregate Economic Variables.} 
Figure~\ref{Figure:learned equilibria open} shows intuitive economic relationships between various aggregate economic quantities.
For example, hours worked increase with wage, and consumption decreases with increasing price.
Such relationships do not exist in the closed economy shown in Figure~\ref{Figure:Learned Equilibria for Closed RBC Models}, because in most cases, the economy is stuck in a ``bad'' equilibrium with very little work, production, or consumption.
In such a setting, many agent actions, e.g., setting tax rates, increasing wages or prices, simply do not have any effect, e.g., increasing labor.

\paragraph{Analyzing Open \RBCmodel{} Solutions.} 
The open \RBCmodel{} model admits a wide spectrum of qualitatively distinct solutions with distinct outcomes (e.g., consumer utility, hours worked, prices, and taxes), and trends that align with economic intuitions.
To study the differences between outcomes, Figure \ref{Figure:Episode Rollout} shows two rollouts sampled from different converged solutions, revealing qualitative similarities along with some distinct behaviors.
\begin{itemize}
    \item In these examples, firms can profit by either focusing on consumers or the export market; they tend to set prices high at first and sell mainly to the export market. Some firms then lower prices halfway through the episode and begin selling to consumers. 
    \item The consumers tend to work in intense ``bursts'', and only consume when prices are lower at the end. Note that each firm receives a non-zero number of hours of labor.
    \item Lower-technology firms, i.e., firms 3, 4, 8, and 9, whose production functions rely more on labor, consistently set high prices and lower wages towards the end, independent of their capital.
    \item Higher-technology firms lower prices over time and keep wages high. This is intuitive, as high technology firms can produce more efficiently and hence face lower labor costs per unit of production.
    \item The firms start with different endowments of capital and invest a fixed percentage of their profits over time, with higher tech firms being more profitable and thus able to sustain higher investment in capital.
\end{itemize}

\section{
    Discussion
}\label{Section:Discussion}
Economic models often assume the existence of a small number of representative agents whose behavior is simple and analytically tractable.
Economists have long understood that these assumptions are unrealistic, and that it could be desirable to define models with more complexity and heterogeneity.
Yet once the models are no longer analytically tractable, finding solutions requires new computational tools.

In this work, we showed how to adapt multi-agent RL to enable meaningful economic analysis of general equilibrium models. 
By using these algorithmic and computational tools, it is possible to weaken modeling assumptions and increase the scale of economic simulations and analysis.

On the other hand, there are also limitations to this approach.
\citet{daskalakis2009complexity} showed that computing equilibria for general-sum games is hard in terms of computational complexity, even for simple matrix games.
There are no theoretical guarantees yet that our framework can find \emph{all} equilibria in sequential economic games.
However, our best-response analysis suggests our framework does discover meaningful local equilibria.

Thus, our hardware-accelerated MARL approach enables economic analysis with a large number of agents allows analyzing complex economic problems in a practical way, while theoretical guarantees are to be developed further.
Hence, our current framework is at least an exploratory tool for finding qualitatively different solutions, e.g., by varying initial conditions, environment parameters, or conditioning sampling, and provides exciting avenues for future research.

\bibliographystyle{ACM-Reference-Format}
\bibliography{references}

\newpage
\appendix

\section{Ethics Statement}
Our work proposes a framework to model economies using Multi-Agent Reinforcement Learning and thus may be used to draw implications about the real world. 
Our findings and used simulations are purely for research purposes and should not be used to make decisions in real-world systems. 
Furthermore, our framework should not be used to explore methods to increase discrimination or unfairness in real-world systems.

\paragraph{Assumptions, limitations, and ethical implications of using ML for economics.}
All choices in the economic simulation model, RL algorithms, reward functions, etc, play an important but difficult-to-understand role in equilibria selection and policy design. As in all ML applications, there are assumptions and limitations in the methodology. This has ethical implications for their use in future policy design applications.

\paragraph{Mitigation strategies and interdisciplinary research.}
Economic simulation enables studying a wide range of economic incentives and their consequences, including models of stakeholder capitalism. However, the version of the simulation as used in this work is not an actual tool that should be used for policy making.

Many design choices influence the eventual policy recommendations. For example, the designer is free to set the social welfare objective that the government optimizes for. As such, it is crucial that these choices are debated and made in a socially acceptable fashion by all stakeholders, and made transparent and accessible to all. 

More generally, to mitigate ethical risk, further mitigation strategies may include performing a what-if analysis over worst-outcomes, opening research results to domain experts (social scientists, ethics experts, etc), and open-sourcing the research results, amongst others. In all, the design and use of ML for policy recommendations will require robust, multilateral discussion and careful consideration of ethical risk, potential harm, and which trade-offs are being made.

We now detail some assumptions, limitations, and potential ethical risk among different dimensions of using ML for economics. \emph{We stress that there can be more (unknown) aspects that we do not address here.} As such, we see this discussion as a starting point of discussion for the ML and economics community.

\paragraph{Economic simulation and data.}
While the current version of the economic simulation provides only a limited representation of the real world, we recognize that future, large-scale iterations can still contain biases and unrealistic assumptions. Furthermore, non-representative simulation environments may result in biased policy recommendations. For instance, the under-representation of communities and segments of the work-force in training data might lead to bias in simulations that build on those and lead to biased AI policies. As such, collecting more representative data is a key challenge for future research in using ML for economic policy recommendations.

Our RBC model is a stylized model of real economies. RBC models are a commonly used class of economic models (see e.g. \citet{smets2007shocks}). However, as any model, it contains assumptions and stylizations. Future simulations may miss (un)known features that pertain, e.g., to equity and equality in the economy. Therefore, using simulations that are not representative or well-calibrated, can exacerbate or create new socio-economic issues.

We list a few salient features and assumptions below, although we cannot exhaustively enumerate all features that may be relevant in future research.

\begin{itemize}
    \item Our RBC model features consumers that differ in skill and perform different amounts of work. However, we do not model more fine-grained distributional features, such as educational attainment, wealth, inheritance, geography, or others. 
    \item Similarly, firms produce a single good only and can invest and pay wages. Any worker can work for any firm. We do not model hiring practices, the geographic location of firms, non-monetary incentives or benefits (e.g., health insurance). To accurately model inequity in the real world, including such features may be necessary. 
    \item On the government and societal level, we model tax policies and simple redistribution of tax revenue. We do not model targeted redistribution, tax credits, application-specific subsidies (e.g., education support). We do not model trading, inflation, debt, and other macroeconomic features that may impact social groups disparately.
\end{itemize}

Our RBC is more general than commonly used models: we do not enforce market clearing, for instance. Market clearing is an unrealistic assumption that supply always meets demand. Economic theory uses such constraints to make analysis tractable. In contrast, our learning approach is flexible and does not require such simplifying assumptions. 
We also assume that all agents can observe the wages offered by all the firms. However, in the real world not all agents have equal access to information -- and this is a feature that can be studied by future research. As such, we view the flexibility of our learning approach as a positive, in that our framework may allow for studying more representative models.

\paragraph{Choice of economic incentives and rewards.}
Agents optimize their behavior given economic incentives, as modeled by their reward function. As such, future economic AI policies should clearly describe for which reward function they were optimized. Furthermore, more research is needed to understand how the choice of reward function influences the resulting policies, and how social and ethical values can be transparently encoded in reward functions. It is also an open question which ethical/social norms and values can or cannot be quantified, and how to encode trade-offs between conflicting values.

For example, the planner optimizes its policy to maximize ``social welfare'', a standard economics concept. However, the definition of social welfare heavily influences the resulting policy and social outcomes. For example, \citet{zheng2020ai} used equality times productivity as their objective and showed the resulting AI income taxes can improve equality over classic tax models. Standard economic works often use the utilitarian objective (sum of all agent rewards). An alternative is the Rawlsian objective (social welfare is the reward of the lowest-income agent). We emphasize the choice of social welfare is flexible and a choice made by the designer(s) and users of the framework. 

Another key example is the discount factor used to weight rewards over time. Whether to emphasize short-term vs long-term rewards is a social choice that has ethical implications. For example, firms may emphasize short-term profits over long-term health issues, which may disparately impact different social groups.

\paragraph{Choice of agent model.}
The behavior of agents is determined by the policy model, e.g., the neural networks used in our work. Neural networks are universal function approximators, given enough width (or depth) in their layers. However, in practice, neural networks may still encode structural biases and only parameterize a particular subspace of all theoretically possible policy models. For our networks, a particular concern might be architectural constraints: our policy networks are not recurrent (so only consider the current state) and sometimes don’t allow correlated actions. With enough parameters these networks are still capable of representing a wide range of policies, but these architectural constraints represent implicit priors which conceivably might not reflect human decision-making. As such, more research is needed on what the limits are of neural networks in terms of emulating human behaviors, and to what extent more and diverse datasets can help alleviate such concerns.

\paragraph{Choice of algorithms and learning strategies.}
The RBC model is an economic “game” that has multiple equilibria. It is not well understood theoretically to which equilibria a given RL algorithm converges. Indeed, previous work has studied MARL beyond independent learners, including Nash-Q \citep{hu2003nash}, WoLF \citep{bowling2002multiagent}, and MADDPG \citep{lowe_multi-agent_2017}. This extends to our use of structured curricula, reward shaping, and other forms of multi-agent learning algorithms or strategies. These methodological choices can all impact the equilibria one finds (or doesn’t find) using ML.

This is important because different equilibria can have different levels of social welfare and granular social outcomes (e.g., equality, type of work performed, unemployment level). From an ethical point of view, it is therefore possible that certain choices of algorithms, etc, may bias policy recommendations and simulation outcomes to socially or ethically undesirable situations. For instance, certain social groups in the simulation may be disparately impacted by policy recommendations. Therefore, it is important for future research to analyze how different RL algorithms may selectively converge to certain equilibria, and how one might enumerate all possible equilibria. This is still a significant theoretical and empirical challenge. 

Defining and justifying the objectives for the social planner and other methodological choices is a complex discussion, and requires a more in-depth understanding of the functioning of ML that is beyond the scope of this work. This requires multilateral, interdisciplinary discussion on, for example, what the preferred social choice is with respect to the definition of social welfare and constraints.

\paragraph{Choice of hyperparameters.}
RL algorithms may converge to different solutions depending on the chosen hyperparameters, e.g., learning rate, entropy regularization, or discount factor. For instance, the level of entropy regularization regulates the exploration-exploitation trade-off in actor-critic methods, a form of on-policy RL as used in our work. It is known that actor-critic methods may get stuck in suboptimal local maxima. This issue may be exacerbated in the multi-agent setting, where there are multiple equilibria, and it is unknown how algorithms converge towards different equilibria. As such, it is possible that certain choices of hyperparameter can encode structural biases towards certain outcomes in the simulation. These potential limitations are an area for future research.

\paragraph{Robustness of Deep RL.} 

A key question is how robust learned policies are to perturbations in the simulation (parameters). This has ethical implications: policies that do well in simulation, may not do well in the real world if, e.g., income distributions differ between sim and real. As such, simulations that are not representative (enough) may lead to policy recommendations that disadvantage underrepresented social groups in the real world.
More generally, it is well-known that deep learning models and RL policies can be very brittle and may not generalize well to unseen environments. As such, more robustness analysis should be done on any policy recommendation that is based on deep RL and related methods.

\paragraph{Explainability and Simplicity of AI policies.}
Even though AI policies may be effective, they may use intricate, unexplainable patterns in their input data to achieve high performance. Moreover, their behavior may vary wildly between different inputs. As a hypothetical example, an RL agent may make significantly different tax rate recommendations for people with slightly different income or education levels. Such behaviors can disproportionally affect underprivileged social groups, and have unintended short/long-term economic consequences, especially if models are not well-calibrated. It is an open question on what level/amount of data, or specific policy constraints, could mitigate such potential risk and harm. We also note that if one wanted to restrict the class of policies to ones that are sufficiently explainable, the same model-free policy optimization scheme could still be applied. Indeed, this is a big potential advantage of the RL approach.
\end{document}